# Quantum dynamics and entanglement of spins on a square lattice

N. B. Christensen[1,2], H. M. Rønnow[3,2], D. F. McMorrow[4,5], A. Harrison[6,7], T. G. Perring[5], M. Enderle[7], R. Coldea[8,9], L. P. Regnault[10], and G. Aeppli[4]

[1] *Materials Research Department, Risø National Laboratory, Technical University of Denmark, DK-4000 Roskilde, Denmark;* [2] *Laboratory for Neutron Scattering, ETH Zurich and Paul Scherrer Institute, CH-5232 Villigen PSI, Switzerland;* [3] *Laboratory for Quantum Magnetism, École Polytechnique Fédérale de Lausanne (EPFL), 1015 Lausanne, Switzerland;* [4] *London Centre for Nanotechnology and Department of Physics and Astronomy, University College London, London WC1H 0AH, United Kingdom;* [5] *ISIS Facility, Rutherford Appleton Laboratory, Chilton, Didcot OX11 0QX, United Kingdom;* [6] *Department of Chemistry, University of Edinburgh, Edinburgh, EH9 3JJ, United Kingdom;* [7] *Institut Laue-Langevin, 38042 Grenoble Cedex 9, France;* [8] *Oxford Physics, Clarendon Laboratory, Parks Road, Oxford OX1 3PU, United Kingdom;* [9] *Department of Physics, University of Bristol, Bristol BS8 1TL, United Kingdom; and* [10] *CEA, Grenoble, Département de Recherche Fondamentale sur la Matière Condensée, 38054 Grenoble Cedex 9, France*

**Abstract:**

**Bulk magnetism in solids is fundamentally quantum mechanical in nature. Yet in many situations, including our everyday encounters with magnetic materials, quantum effects are masked, and it often suffices to think of magnetism in terms of the interaction between classical dipole moments. Whereas this intuition generally holds for ferromagnets, even as the size of the magnetic moment is reduced to that of a single electron spin (the quantum limit), it breaks down spectacularly for antiferromagnets, particularly in low dimensions. Considerable theoretical and experimental progress has been made in understanding quantum effects in one-dimensional quantum antiferromagnets, but a complete experimental description of even simple two-dimensional antiferromagnets is lacking. Here we describe a comprehensive set of neutron scattering measurements that reveal a non-spin-wave continuum and strong quantum effects, suggesting entanglement of spins at short distances in the simplest of all two-dimensional quantum antiferromagnets, the square lattice Heisenberg system.**

One of the most fundamental exercises in quantum mechanics is to consider a pair of $S = 1/2$ spins with an interaction $J$ between them that favors either parallel (ferromagnetic) or antiparallel (antiferromagnetic) alignment. The former results in a spin $S_{tot} = 1$ ground state, which is a degenerate triplet. Two of the states in this triplet are the possible classical ground states $|\uparrow\uparrow\rangle$ and $|\downarrow\downarrow\rangle$, whereas the third is the coherent symmetric superposition $|\uparrow\downarrow\rangle + |\downarrow\uparrow\rangle$, which has no classical analogue. Even more interesting is antiferromagnetic $J$, for which the ground state is the entirely nonclassical $S_{tot} = 0$ singlet $|0\rangle = |\uparrow\downarrow\rangle - |\downarrow\uparrow\rangle$ consisting of the antisymmetric coherent superposition of the two classical ground states of the pair. The state $|0\rangle$ is an example of maximal entanglement, i.e., a wavefunction for two coupled systems that cannot be written as the product of eigenfunctions for the two separate systems, which in this case are of course the two spins considered individually.

The consideration of spin pairs already suggests a very clear difference between ferromagnets and antiferromagnets, in the sense that in ferromagnets the classical ground states are stable upon the introduction of quantum mechanics, whereas those of antiferromagnets are not obviously so. Indeed, as recently as the midpoint of the last century, it was felt that ordered antiferromagnets might not exist. Subsequent experiment and theory have proven this wrong for most practical purposes. However, recent developments in areas as disparate as mathematical physics and materials science, as well as interest in harnessing entanglement as a resource for quantum computation (1), have led to a renaissance in investigations of purely quantum mechanical effects in antiferromagnets. An early milestone in this field was the realization that antiferromagnetic chains in general do not have ground states with classical order (2), and indeed have excited state spectra that do not correspond to those of the classical systems either (3–5). Another was the observation by Anderson (6) that such nonclassical ground states have remarkable similarities to the coherent quantum wavefunctions describing superconductors—something that is intuitively appealing when it is recalled that the spin wavefunction of the underlying Cooper pairs of electrons correspond to the ground state of the antiferromagnetic spin pair. He suggested that a highly entangled ground state, referred to as ''resonating valence bond'' (RVB) (7), for a square lattice of antiferromagnetically coupled copper ions might be the precursor of the high temperature superconductivity found in the layered cuprates.

However, studies of the insulating parents of the high-temperature superconductors have revealed essentially conventional classical magnetic order in these two-dimensional $S = 1/2$ Heisenberg antiferromagnets (8). In addition, the magnetic excitations (9) and temperature dependence of the magnetic correlations (10–12) are given to remarkable accuracy by classically based theory (13–15). One interesting fact, though, is that the static moment, reflecting the extent to which the true ground state resembles the Néel state |N>, is only ≈60% of that anticipated classically (13, 16). In addition, the ground-state energy is lower than that calculated for |N> (16). Consequently, there are substantial quantum corrections to the ground-state wavefunction, and there can be a certain level of entanglement among spins even on the square lattice (see ref. 17 for the particular example of the $S = 1/2$, XYX model), although to date there has been no direct experimental evidence for this. Indeed, so far, essentially all experimental studies have focused on renormalization of classical parameters rather than on qualitatively new quantum phenomena. The only exceptions have been two papers reporting anomalous zone boundary spin-wave dispersion (18, 19) and a very brief conference report (20). The purpose of the present paper is to report on a definitive inelastic neutron scattering study showing large qualitative effects of quantum corrections for a particular realization of the square lattice $S = 1/2$ Heisenberg antiferromagnet. Furthermore, we draw direct attention to the power of this technique as a diagnostic of entanglement in magnetic systems.

Neutron scattering provides images of the wave-vector, $\mathbf{Q}$, and energy, $E$, dependent magnetic fluctuations in solids (21). The technique reveals the underlying wavefunctions via Fermi's golden rule, according to which the cross-section defining the probability of scattering into a particular direction $d\Omega$ and energy range $dE$ is of the form

$$\left(\frac{d^2\sigma}{d\Omega dE}\right) \sim \left|\langle f|S_{\mathbf{Q}}|i\rangle\right|^2 \quad (1)$$

The operator $S_{\mathbf{Q}}$ appearing in the matrix element between initial and final states $|i\rangle$ and $|f\rangle$ is the Fourier transform $\Sigma_j S_j \exp(i\mathbf{Q}\cdot\mathbf{r}_j)$ of spin operators $S_j$ on sites $\mathbf{r}_j$. A dramatic and pertinent illustration of how neutrons can probe entanglement is shown in Fig. 1 for an isolated square plaquette of antiferromagnetically coupled spins. The $S_{\text{tot}} = 0$ ground state is strongly entangled, being a coherent superposition of the two classical ground states and more energetic configurations containing spins parallel to their neighbors. The operators $S_{\mathbf{Q}}$ connect this ground state to three excited $S_{\text{tot}} = 1$ states (22). It is possible to see immediately, as illustrated in Fig. 1, that the neutron scattering cross-section for the quantum plaquette is radically different around the $(\pi, 0)$ and $(0, \pi)$ points, from that of the classical plaquette, with a single peak at $(\pi, \pi)$, the vector describing the antiferromagnetic order (see *Materials and Methods* for a definition of the notation used to specify wavevectors). For the square lattice antiferromagnet, which is an infinite array of plaquettes, the ground state is the Néel state with quantum corrections and has long

been suspected (23) to have a finite overlap with wavefunctions of the RVB type shown in Fig. 1b for a single plaquette. Correspondingly, the excited states are what have become of the classical spin waves, also with quantum corrections.

The real two-dimensional Heisenberg antiferromagnet considered here is copper deuteroformate tetradeurate (CFTD), which crystallizes from aqueous solution as a stack of nearly square lattices of copper atoms alternating with water layers. The transparent, insulating material has a bluish hue, with the copper ions carrying spin $S = 1/2$. CFTD was chosen because the electrons are strongly localized, implying a negligible role for charge fluctuations, very unlike what is seen in the black insulating parents of the high-temperature superconductors (9). Orbital overlap through intervening formate groups couple neighboring copper spins by antiferromagnetic exchange interactions of a convenient magnitude ($J = 6.19$ meV, equivalent to a temperature of 72 K) for magnetic spectroscopies (19, 24).

**Results**

Neutron time-of-flight spectroscopy with position-sensitive detection allows the magnetic dynamics to be surveyed over large volumes of **Q**-$E$ space simultaneously. The easiest method for displaying these data for a two-dimensional magnet such as CFTD is to plot the intensities as a function of the two-dimensional wavevector for fixed neutron energy loss $\hbar\omega$. The dominant features in such cuts through the dispersion surface (Fig. 2a) are rings, centered on the reciprocal lattice point for the ordered spins. Fig. 2b shows three such images for CFTD. The rings arise from well defined magnetic excitations, known as magnons or spin waves in a classical description. Of particular interest is the behavior near the maximum of the single-magnon dispersion, since anomalies have been reported in the energy of the spin waves along the zone boundary (dashed red lines in Fig. 2) in a number of systems (9, 18, 19). The images in Fig. 2 c and d display the intensity distributions for energies well below and in the vicinity of the zone boundary energy in CFTD. When we compare these results with simulations from linear spin-wave theory in the same energy ranges (Fig. 2 e and f), it is clear that the data at intermediate energies follow the uniform intensity distribution expected (Fig. 2 *a* and *e*), whereas a large modulation in intensity is apparent along the zone boundary, with "holes" around the ($\pi$, 0) points, enclosed in dashed blue squares in Fig. 2*d*. These holes represent large deviations from linear spin-wave theory. They result from quantum mechanical interference and evidence entanglement of nearest-neighbor spins. However, before such a dramatic conclusion is justified, further work and detailed analysis must be undertaken to rule out the possibility that the holes might be an artifact of the dispersion in the peak position.

Fig. 3 *c–f* summarizes the analysis of the single-magnon scattering along the major symmetry directions of the Brillouin zone of the square lattice, and it compares the results with spin-wave calculations (13, 15) indicated by red lines. Constant wavevector cuts at ($\pi/2$, $\pi/2$) and ($\pi$, 0) are shown in Fig. 3 *a* and *b*. Along the M-$\Gamma$ direction we find that a good fit of the single-magnon energies (Fig. 3 *c* and *d*) to the linear spin wave (LSW) theory result is achieved with $J = 6.19(2)$ meV. On approaching the X point, the energy is depressed by 7(1)% relative to the point ($\pi/2$, $\pi/2$), confirming the value of 6(1)% found in our previous lower-resolution study (19). This effect is not expected classically (13, 15) but agrees well with estimates obtained by exact diagonalization (19), series expansions around the Ising limit (25) (blue lines), and Quantum Monte Carlo computations (26) (blue squares).

Fig. 3*e* shows the momentum-dependent spin-wave intensity. At first sight, the main feature is the divergence as the magnetic zone center ($\pi$, $\pi$) is approached, which is in agreement with classical spin-wave calculations (13). By normalizing to the classical predictions for the intensity, we bring out very clearly a spectacular result of our experiments (Fig. 3*f*), namely that along with the relatively modest dispersion of the spin waves along the zone boundary, there is a much stronger intensity anomaly, actually removing 54(15)% of the classically predicted mode intensity at ($\pi$, 0). Thus, there are very large quantum corrections to the wavefunctions when we look at short distances, a result completely at odds with the standard notion of a momentum-independent renormalization (15). Note that to lowest order in spin-wave theory, next-nearest-

neighbor interactions do not lead to any intensity variation along the zone boundary. The intensity anomaly has a half-width in momentum space along the X-M direction of 0.1 Å$^{-1}$, implying that the underlying physical phenomenon has a length scale of 10 Å, or roughly twice the nearest-neighbor Cu-Cu distance. In addition, the wavefunction corrections involve spin correlations with wavevectors of type ($\pi$, 0). Such corrections at this wavevector are exactly what might naively be expected for an RVB state where nearest-neighbor valence bonds entangle spins along the edges of the square lattice— as already occurs for the quantum plaquette in Fig. 1b—but not along the diagonals. This intuitive argument is supported by calculations of mode softening and intensity reduction at the ($\pi$, 0)-type points—reproduced by the green lines in Fig. 3 c–f—for a variational state mixing RVB and Néel order (27, 28). While the variational calculation exaggerates the qualitative features that we have found, most notably yielding a deeper spin-wave energy minimum at ($\pi$, 0), a recent Quantum Monte Carlo study (26) gives zone boundary dispersion and intensity suppression in good agreement with the values found in our experiment.

The ground-state wavefunction |$\Psi$> of the two-dimensional $S = 1/2$ Heisenberg antiferromagnet can be viewed as the sum of the Néel state |N> and correction terms that are responsible for the reduction of the ordered moment. In a spin-wave description, these would be a series of single and multiple spin flips that in turn can be expressed as the Fourier transforms of single- and multiple-magnon excitations. In a neutron scattering experiment, the application of the operator $S_\mathbf{Q}$ to such correction terms leads to additional continuum scattering, usually called "multimagnon" because the momentum and energy imparted by the neutron to the sample can be shared among pairs of classical spin waves. If we move away from simply including single spin flip corrections to |$\Psi$> and also take account of RVB-like terms (23), as illustrated in Fig. 1e, the structure of the continuum can be substantially altered from that given by two-magnon calculations, as borne out by the different predictions given by competing theoretical scenarios (26, 29).

The unambiguous observation of multimagnon continuum scattering is highly nontrivial, because of the difficulty of isolating a weak, diffuse signal from the one-magnon signal in the presence of scattering from phonons, nonuniform background, etc. One particular setting of the MAPS spectrometer was found to provide a "clean" window within which to search for the existence of continuum scattering, namely for energies in the range $8.7 < \hbar\omega < 11.5$ meV with the two-dimensional Cu layers of CFTD aligned parallel to the incident beam. Multimagnon processes (29) are expected to produce a continuum inside the dispersion cone (Fig. 2a). The data in Fig. 4a reveal extra intensity in this region, an effect that is made even clearer when integrating over a range of $L$ (Fig. 4 b and c). Indeed, although the single-magnon model (Fig. 4 b and c, red lines) provides a remarkably good description of the data, including those for the spin-wave cone emanating from the structural zone centers at $|H| = 1$, it fails to account for the observed filling of the cone. It is only by including a two-magnon contribution in the model (Fig. 4 b and c, blue lines) that a satisfactory account of the data is achieved.

Definite proof that the additional scattering near $(\pi, \pi)$ in Fig. 4 is magnetic comes from polarized neutron scattering. This technique allows separation, in a model-independent way, of the fluctuations transverse and longitudinal to the ordered moment that, to lowest order in perturbation theory, are dominated by single- and two-magnon events, respectively. Fig. 5 displays data at intermediate energies and wavelengths, demonstrating clearly that the filling of the spin-wave cone seen by using MAPS arises from the longitudinal response that can only be associated with multimagnon processes. The intensity ratio of two-magnon to single-magnon scattering is determined by the reduction $\Delta S$ of the ordered moment from its classical value of 1/2. Quantitative analysis of our data yields $\Delta S = 0.23 \pm 0.02$, in accord with the theoretical estimate $\Delta S = 0.197$ (ref. 13).

Of great interest—in light of the anomalies found in the single-magnon response (Fig. 3)—is the continuum at the zone boundary, shown in Fig. 6. The high-energy polarized data for the transverse component (Fig. 6 a and b) display the same feature as that found using MAPS: The peak response at the $(\pi, 0)$ point is depressed in energy relative to $(\pi/2, \pi/2)$ and is also less intense. The latter effect appears less dramatic in Fig. 6 a and b than in Fig. 3f because of the poorer momentum resolution of the polarized beam instrument, which averages over the dip at (π, 0), as well as the fact that the unpolarized MAPS data are simultaneously sensitive to the longitudinal mode, which appears even more suppressed at (π, 0) relative to (π/2, π/2) than the transverse mode (Fig. 6 *c* and *d*). In addition to the peak at the renormalized spin-wave energy, weak but significant intensity extends to energies above the main spin-wave peak in both the transverse (Fig. 6 *a* and *b*) and longitudinal channels (Fig. 6 *c* and *d*). Such intensity must be present to balance the total moment sum rule, which states that $S(S + 1)$ is the sum of contributions from the ordered moment, the spin waves and the continuum (30).

Quantum Monte Carlo calculations (26), indicated by dashed green lines in Fig. 6, broadly account for the data in both channels. Most notably, there is a pile-up of intensity in the longitudinal channel at energies just above the sharp peak in the transverse response. This is in contrast to the two-magnon theory, whose results are also shown in Fig. 6 (solid red and blue lines), which posits a less pronounced maximum in the longitudinal response. In addition, the two-magnon theory does not explain the momentum dependence of the amplitude of the longitudinal response, just as conventional spin-wave theory does not account for the amplitudes of the transverse response at the zone boundary. An alternative view is provided by the variational RVB approach (27, 28), where the continuum above the zone boundary arises from fermionic (spinon) excitations. Unfortunately, no predictions have been made for the longitudinal continuum. Clearly, further refinements to experiment and theory are required before a definitive conclusion can be drawn for the nature of the continuum we have observed in our experiments.

**Summary and Conclusions**

The square lattice $S = 1/2$ Heisenberg antiferromagnet has long been regarded as having only minor quantum corrections to its behavior (14, 31). We have shown that at short distances, this is actually not the case, and that there are large momentum-dependent corrections to the spin waves. In addition, we have provided the first observation of the multimagnon continuum in this simplest of two-dimensional quantum magnets. As for the spin waves, the most significant discrepancy between our results and conventional theory appears at the zone boundary. Quantum Monte Carlo accounts well for all of the data, although it does not give any particular physical insight into why classical theory fails. Conventional spin-wave and two-magnon calculations give an excellent quantitative description of the long and intermediate wavelength fluctuations. At short wavelengths, a variational wavefunction that explicitly mixes RVB and classical (Néel) terms provides a qualitative, but not quantitative, account of the intensity anomaly at ($\pi$, 0), suggesting that the experiments as well as the Quantum Monte Carlo data are due to a significant admixture of RVB terms in the ground-state wavefunction. Indeed, what they show is that the matrix elements (Eq. **1**) between the corrected ground state and the magnon eigenstate actually contain destructive interference terms between valence bond corrections and the classical states. Specifically, these arise because of the nonorthogonality of the Néel state |N> and the RVB corrections; within the two-magnon theory, such interference terms do not occur because the single-magnon states in the expansion for the ground state are orthogonal to |N>. Fig. 2*d* is an image of these interference effects (as for the plaquettes illustrated in Fig. 1), which are most severe near the zone boundary point ($\pi$, 0), indicating the entanglement of spins on adjacent sites of this exceptionally simple magnet without disorder or geometrical frustration. In fact, nearest-neighbor singlet corrections to the ground-state wavefunction are manifested independently and directly in the bond-energy, which is +*J*/4 for the triplet and –3*J*/4 for the singlet, whereas the average energy of uncorrelated bonds is zero. Because the $S = 1/2$, square lattice Heisenberg antiferromagnet has a ground-state energy per bond of $E_b \approx -0.34J$ (23), which is lower than the –0.25*J* of the classical Néel state, the corrections must overrepresent singlets compared with random quantum disorder. It is worth remarking that, in the $S = 5/2$ square lattice system $Rb_2MnF_4$ the zone boundary spin waves are dispersionless to within 1%, and only a weak continuum of two-magnon states has been observed (32).

In summary, we have performed precision experiments using state-of-the art neutron scattering techniques to provide a complete description of the spin dynamics of the two-dimensional square lattice Heisenberg antiferromagnet—at all relevant wavevectors and energies in the one- and multimagnon channels. The quantum nature of the dynamics in this system is shown to manifest itself in the one-magnon channel as a mild zone boundary dispersion in the energy, and a much larger dispersion in the intensity, whereas the continuum of multimagnon events is imaged directly for the first time and has, as the total moment sum rule suggests, an intensity determined purely by the reduction in the size of the ordered moment because of quantum fluctuations. We compare our data with a variety of analytical theories and computer simulations, including expansion from the Ising limit, flux-phase RVB and Quantum Monte Carlo to show that no currently available analytical theory accounts for the short-wavelength dynamics probed at the zone boundary.

Our discoveries have implications not only for quantum magnets as laboratories for the study of entangled states (17, 33–35), but also for one of the principal problems of modern condensed matter physics, namely that of the high-temperature superconductors. The high-$T_c$ cuprates are doped two-dimensional Heisenberg quantum antiferromagnets and many believe that this is the key to a full quantitative description of the cuprates. In the parent compounds of the cuprates, Raman scattering (36) and infrared absorption measurements (37) have revealed intense, broad responses extending up to several times the zone boundary magnon energy. Multimagnon processes, treated in linear spin-wave theory, have emerged as essential to account for these observations (38, 39), but despite some successes these models do not account for the full range of experimental anomalies. It has been suggested (39) that the discrepancies between the optical data and the multimagnon theories are due to strong local deviations from the Néel state, exactly what is seen directly by our neutron scattering measurements. Although the role of charge fluctuations cannot be neglected in the cuprates, the discovery of nearest-neighbor quantum effects already present in the Heisenberg antiferromagnet supports resurgent theoretical ideas (40) about the importance of RVB-type correlations that become dominant upon the introduction of frustration or mobile charge carriers, leading to the destruction of Néel order.

**Materials and Methods**

Measurements were performed on single crystals of CFTD [Cu(DCOO)$_2$·4D$_2$O], grown by slow evaporation in a vacuum dessicator at ≈25°C of a Cu(DCOO)$_2$ solution, which was prepared by dissolving the deuterated carbonate, Cu$_2$CO$_3$·(OD)$_2$·D$_2$O, in a solution of d2-formic acid in D$_2$O (41). The carbonate was synthesized by adding Na$_2$CO$_3$ to a solution of Cu$^{2+}$ in D$_2$O.

CFTD has a monoclinic low-temperature crystal structure, space group $P2_1/n$ with $a = 8.113$ Å, $b = 8.119$ Å, $c = 12.45$ Å, and $\beta = 100.79°$. The Cu$^{2+}$ ions form nearly square lattice planes parallel to the $a$ and $b$ axes separated by layers of crystal-bound heavy water. Each Cu$^{2+}$ ion carries a localized $S = 1/2$ spin that interacts with its four in-plane nearest neighbors located ≈5.739 Å away through antiferromagnetic Heisenberg exchange interactions $J$ mediated by intervening formate groups, whereas interplane interactions $J_c$ are much weaker, $J_c \approx 10^{-5}$–$10^{-4} J$. Quantum chemistry implies that further neighbor interactions within the planes of this ionic salt are also negligible, consistent with the perfect agreement of the measured spin-wave energies with the numerical results of refs. 25 and 26, which employ nearest-neighbor coupling only and would disagree with the measured dispersion if further neighbor interactions were included. The basic Hamiltonian of the system is therefore $H = J \Sigma_{<il>} S_i \cdot S_l$, where the summation extends over in-plane nearest-neighbor pairs only.

In momentum space, the reciprocal crystallographic lattice is spanned by the reciprocal lattice vectors $a^*$, $b^*$, and $c^*$ and indexed $\mathbf{Q} = (H, K, L)$. Below an ordering temperature of $T_N = 16.5$ K, CFTD forms a simple antiferromagnetic structure where neighboring spins align anticollinearly almost along the $a^*$ direction (42), leading to magnetic Bragg scattering at $\mathbf{Q} = (1, 0, 1)$ and equivalent positions, which correspond to the antiferromagnetic zone center $\mathbf{Q}_{2D} = (\pi, \pi)$ in the reciprocal lattice of the plaquette and of the two-dimensional square lattice model. One can transform from $\mathbf{Q} = (H, K, L)$ to $\mathbf{Q}_{2D} = (Q_x, Q_y) = \pi(H + K, H - K)$, while the L component of the scattering vector is irrelevant to the pure two-dimensional model. Throughout this article, we use the two-dimensional square lattice notation, supplemented with crystallographic (H, K, L) notation in Figs. 2, 4, and 5.

We would like to acknowledge helpful discussions with P. Lee, K. Lefmann, D. A. Tennant, R. R. P. Singh, A. Sandvik, A. Läuchli and R. A. Cowley. Work in London was supported by Research Councils United Kingdom Basic Technologies and Royal Society-Wolfson Fellowship programs.

**Fig. 1.** From classical and quantum plaquettes to the square lattice. (*a*) Classical ground states for a system of four $S = 1/2$ spins on the vertices of a plaquette. (*b*) The quantum mechanical ground state for a plaquette corresponds to the sum of two terms, each consisting of two singlet pairs along parallel edges of the plaquette. Here the ground state of a single dimer is $|-\rangle = |\uparrow\downarrow\rangle - |\downarrow\uparrow\rangle$ and the ground state of the plaquette is as follows:

$$\left|\begin{array}{c}-\\-\end{array}\right\rangle + \left|\;|\;|\;\right\rangle = -2\left|\begin{array}{c}\uparrow\downarrow\\\downarrow\uparrow\end{array}\right\rangle - 2\left|\begin{array}{c}\downarrow\uparrow\\\uparrow\downarrow\end{array}\right\rangle + \left|\begin{array}{c}\uparrow\uparrow\\\downarrow\downarrow\end{array}\right\rangle + \left|\begin{array}{c}\uparrow\downarrow\\\uparrow\downarrow\end{array}\right\rangle + \left|\begin{array}{c}\downarrow\downarrow\\\uparrow\uparrow\end{array}\right\rangle + \left|\begin{array}{c}\downarrow\uparrow\\\downarrow\uparrow\end{array}\right\rangle$$

(For simplicity, we have neglected normalization factors throughout this text.) (*c*) The neutron scattering structure factor $S(\mathbf{Q})$, obtained by integrating Eq. **1** over all energies, for the classical system in *a* is simply given by $S(\mathbf{Q}) = |1 - \exp(iQ_x) - \exp(iQ_y) + \exp(i(Q_x + Q_y))|^2 = \sin^2(Q_x/2)\sin^2(Q_y/2)$. (*d*) The full quantum mechanical structure factor can be evaluated by summing over the $S_{tot} = 1$ excited states (22) to obtain $S(\mathbf{Q}) = 1 - \cos^2(Q_x/2)\cos^2(Q_y/2)$. (*e*) Portion of the infinite square lattice ground state, viewed as the Néel state (weight $\eta$) with quantum corrections (weights $\nu_n$), of which one resonating valence bond configuration is sketched.

**Fig. 2.** Imaging the single-magnon excitations. (*a*) Spin-wave dispersion surface $\hbar\omega_\mathbf{q} = 2J_{eff}(1 - \gamma_\mathbf{q}^2)^{1/2}$ for the $S = 1/2$, square lattice Heisenberg antiferromagnet as predicted by linear spin-wave theory (13, 15), which assumes a classically ordered (Néel-type) ground state. Here $\gamma_\mathbf{q} = (\cos(Q_x) + \cos(Q_y))/2$ and $J_{eff} = 1.18J$ is an effective exchange interaction, renormalized by the nonclassical contributions to the ground state (15). The color scale indicates the expected intensities $I_\mathbf{q}$ that by Eq. **1** are proportional to squared matrix elements connecting the ground and excited states. In linear spin-wave theory (13) $I_\mathbf{q} \propto ((1-\gamma_\mathbf{q})/(1 + \gamma_\mathbf{q}))^{1/2}$. (*b*) Data produced by averaging over four equivalent Brillouin zones. The raw data were obtained from the 16 m$^2$ detector bank of the MAPS spectrometer at ISIS, Rutherford Appleton Laboratory with the two-dimensional Cu layers of CFTD aligned perpendicular to an incident neutron beam with $E_i = 36.25$ meV. The false color scale represents the measured neutron scattering cross-section in constant

energy slices (as indicated by the shaded plane in *a*) through the dispersion surface. (*c* and *d*) Constant energy slices measured at 8–10 meV and 14–15 meV, plotted with the horizontal *x* and *y* axes labeled in the reciprocal space of CFTD where (1, 0, L) and (0, 1, L) correspond to (π, π), as indicated by the diagonal axes in *c*. The out-of-plane component to the scattering vector is determined by the kinematical constraints and was L = 1.25 and L = 2.0 in *c* and *d*, respectively. Dashed blue squares in *d* show the location of reciprocal space points equivalent to (π, 0). (*e* and *f*) Simulation of the scattering expected in linear spin-wave theory in the energy ranges shown in *c* and *d*. The calculations include resolution convolution as well as a dipolar factor in the cross-section for magnetic scattering of neutrons (21) that leads to an intensity modulation depending on the angle between the scattering vector **Q** and the direction of the ordered Cu moments. The dashed red lines in *a*, *b*, and *e* indicate the magnetic Brillouin zone boundary.

**Fig. 3.** Summary of fitting the single-magnon excitations. (*a* and *b*) Constant wavevector cuts at the high-symmetry points (π/2, π/2) and (π, 0) of the magnetic Brillouin zone boundary (Fig. 2*a*). By fitting the data, the position (red arrows) and intensity of the spin excitation is extracted, allowing the dispersion relation and corresponding intensity variation along the main symmetry directions of the two-dimensional Brillouin zone (indicated by solid lines in the *Inset* in *e*) to be derived. (*c*) Full dispersion relation for the dominant single-magnon excitations. (*d*) Dispersion divided by the linear spin-wave prediction (Fig. 2*a*) with $J$ = 6.19 meV. (*e*) Intensity of the spin excitations along the same momentum space path as in *c*. (*f*) Intensity divided by the linear spin-wave prediction that best describes the data between (π, π) and (2π, 0). Red, blue, and green lines in *c*–*f* indicate the dispersions and intensities predicted by linear spin-wave theory (13, 15), series expansion techniques (25), and flux phase RVB (27), respectively. The solid blue squares are results of a Quantum Monte Carlo computation (26) of the excitation spectra at (π/2, π/2) and (π, 0). Note that had we chosen to employ a nonlinear background model in *a* and *b*, we would have obtained slightly larger statistical errors on the zone boundary intensities, but the significant spectral weight anomaly at (π, 0) would remain, as proven unambiguously by its presence in the lower resolution polarized data shown in Fig. 6.

**Fig. 4.** Imaging the two-magnon excitations. (*a*) Raw data from the MAPS spectrometer with the sample oriented such that the incident neutron beam ($E_i$ = 24.93 meV) is parallel to the two-dimensional Cu layers of CFTD. The third component to the scattering vector K = –1.07 is fixed by the kinematical constraints. The extremely weak coupling between the layers in the *c* direction ($J_c \approx 10^{-5}$–$10^{-4}$) implies that the excitations are independent of the reciprocal space coordinate L. In this case, at a given energy transfer, the circles shown in Fig. 2*b* become cylinders along the L direction. The intercept of the cylinder with the detector trajectories in **Q** and energy produces parabolic cuts through the dispersion surface. In the region 0.3 < L < 1, sharp, single-magnon modes run approximately parallel to the L direction. At higher values of L, the detector trajectories produce a filling-in due to single-magnon events. (*b* and *c*) Intensity integrated over the region indicated by dashed lines in *a*. *b* is a magnified version of *c*. The intensity between the single-magnon modes is evidently higher than in the regions immediately to their exterior. This additional intensity arises from multimagnon processes. The solid red line represents the calculated single-magnon response convolved with the full instrumental resolution and gives a poor account of the data between the peaks. The solid blue line is the sum of the calculated one and two-magnon contributions, calculated by using Monte Carlo sampling (29). The relative spectral weight of the two contributions is completely specified by the quantum mechanical reduction $\Delta S$ of the staggered ordered moment <*S*> from its classical value of 1/2. The red line corresponds to the classical case $\Delta S = 0$ with no multimagnon continuum.

**Fig. 5.** Long-wavelength spin excitations. False color maps of the intensity of the low-energy spectrum of single-magnon excitations polarized transverse to the ordered moment (*a*) and of multimagnon excitations polarized along the ordered moment (*b*), measured at **Q** = (H, 1, 0). Solid white lines indicate the prediction of linear spin-wave theory (15) with J = 6.19 meV. (*c–f*) Raw data and fits upon which the color images in *a* and *b* are built. (*c* and *d*) Transverse and longitudinal excitations at 5 meV. (*e* and *f*)

Transverse and longitudinal excitations at 9 meV. The solid lines in *c–f* are fits to semiclassical theory for one- and two-magnon excitations, calculated by using Monte Carlo sampling (29) with the value $\Delta S = 0.23$ for the reduction in the staggered moment taken from the time-of-flight experiment (Fig. 4). With $J = 6.19$ meV fixed this leaves only a single overall intensity scale factor to be adjusted. The data were obtained at a sample temperature of 1.5 K by using polarized neutron scattering ($k_f = 2.662$ Å$^{-1}$) at the IN20 spectrometer at the Institut Laue- Langevin (Grenoble, France).

**Fig. 6.** Short wavelength spin excitations. The transverse and longitudinal responses at the zone boundary points ($\pi/2$, $\pi/2$) (*a* and *c*) and ($\pi$, 0) (*b* and *d*) measured at, respectively, **Q** = (0, 5/2, 0) and **Q** = (1/2, 5/2, 0) at a temperature of 1.9 K. Note that the broad **Q**-resolution of IN22 relative to the MAPS spectrometer partially washes out the rather narrow dip in intensity at ($\pi$, 0) evident in Fig. 3*f*. Solid lines are single-magnon (red in *a* and *b*) and two-magnon (blue in *c* and *d*) responses from linear spin-wave theory ($J = 6.19$ meV) convolved with the full instrumental resolution. The value $\Delta S = 0.23$ obtained from the time-of-flight experiment (Fig. 4) was used to fix the relative intensities in the transverse and longitudinal channels, leaving only one overall scale factor to be adjusted. Dashed lines are Quantum Monte Carlo results (26) convolved with the instrumental energy resolution. In *a* and *b* the sharp single-magnon and transverse continuum components (26) are plotted separately. The data were obtained by using polarized neutron scattering ($k_f = 2.662$ Å$^{-1}$) at the IN22 spectrometer at the Institut Laue-Langevin (Grenoble, France).

# Figure 1

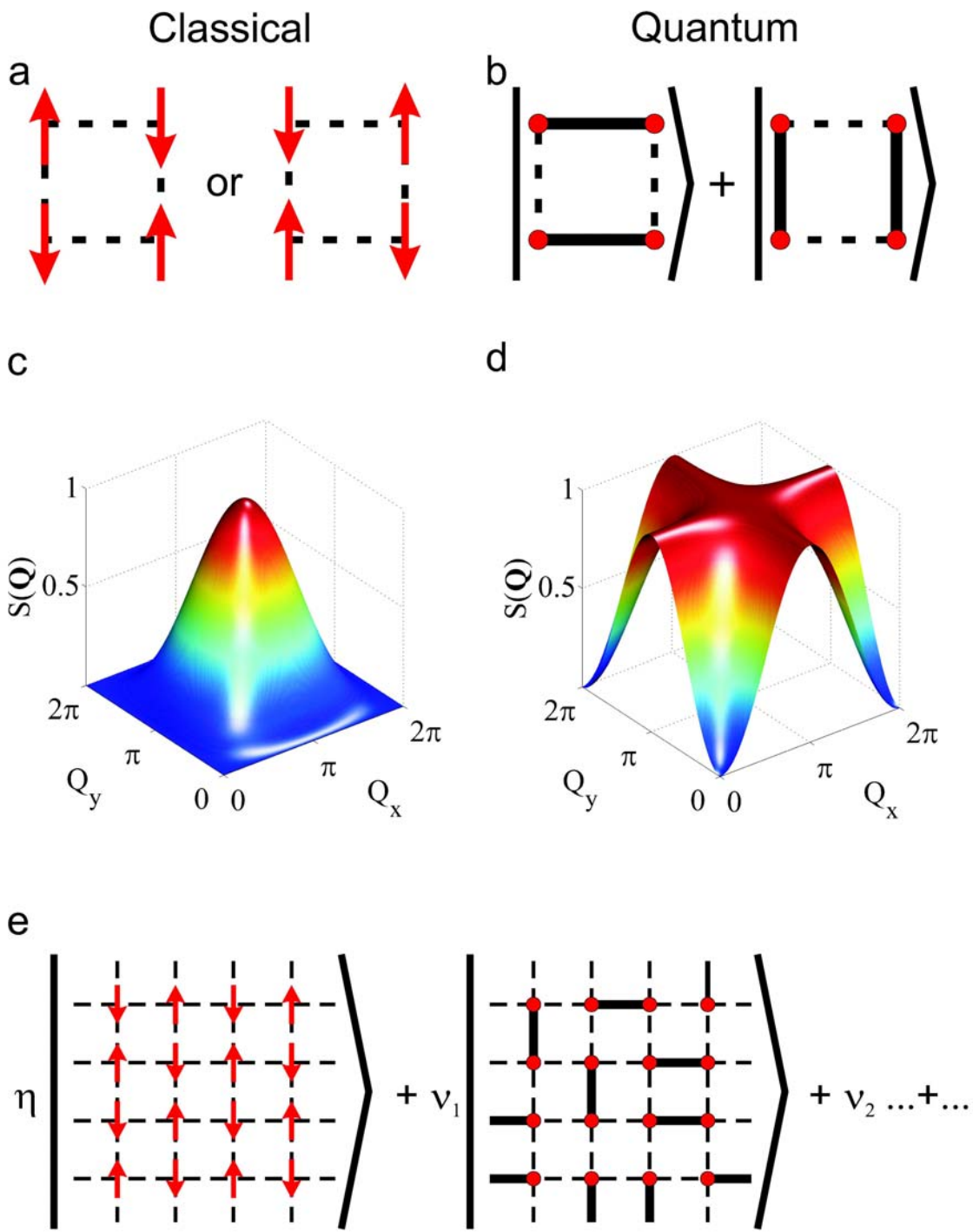

# Figure 2

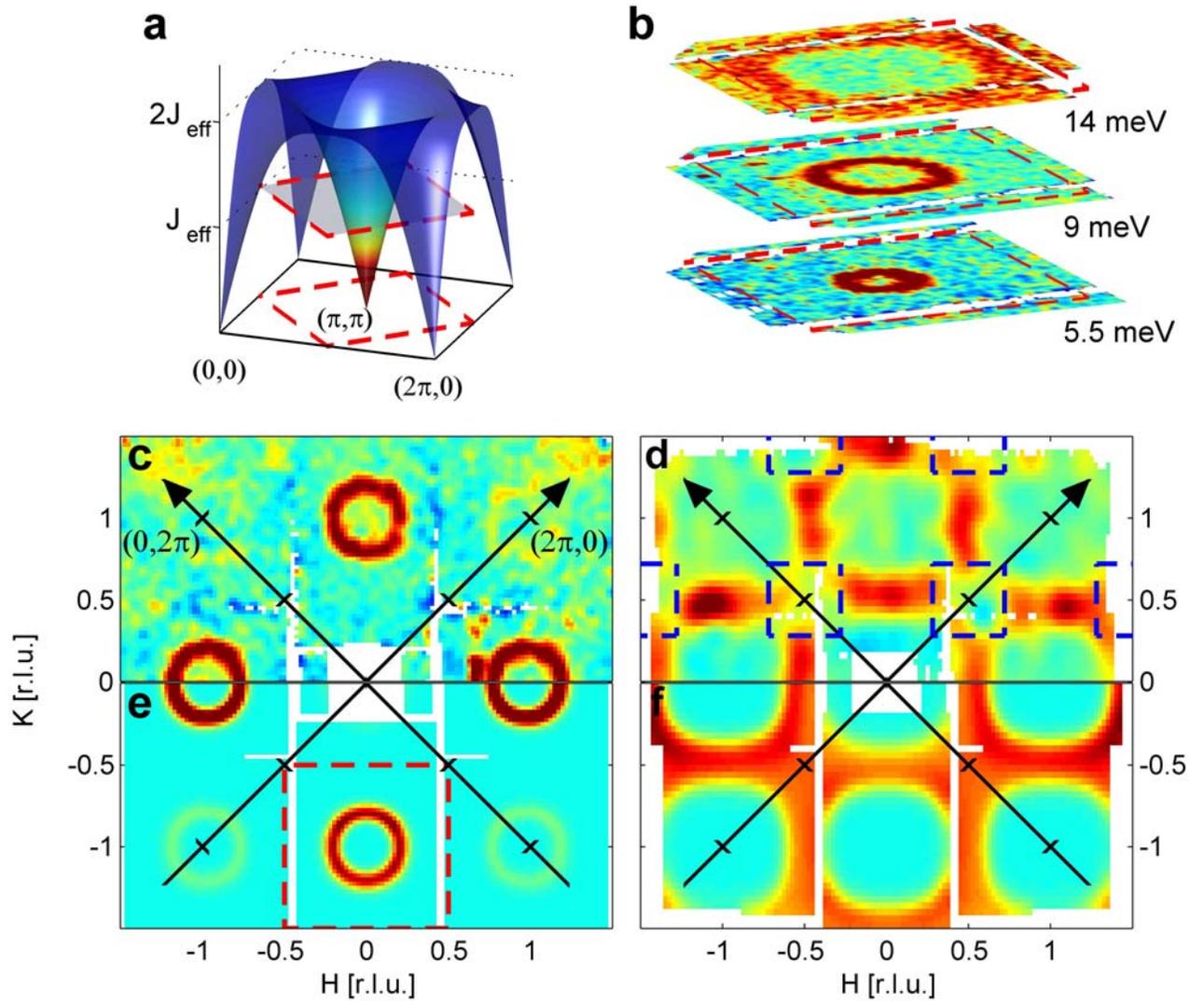

# Figure 3

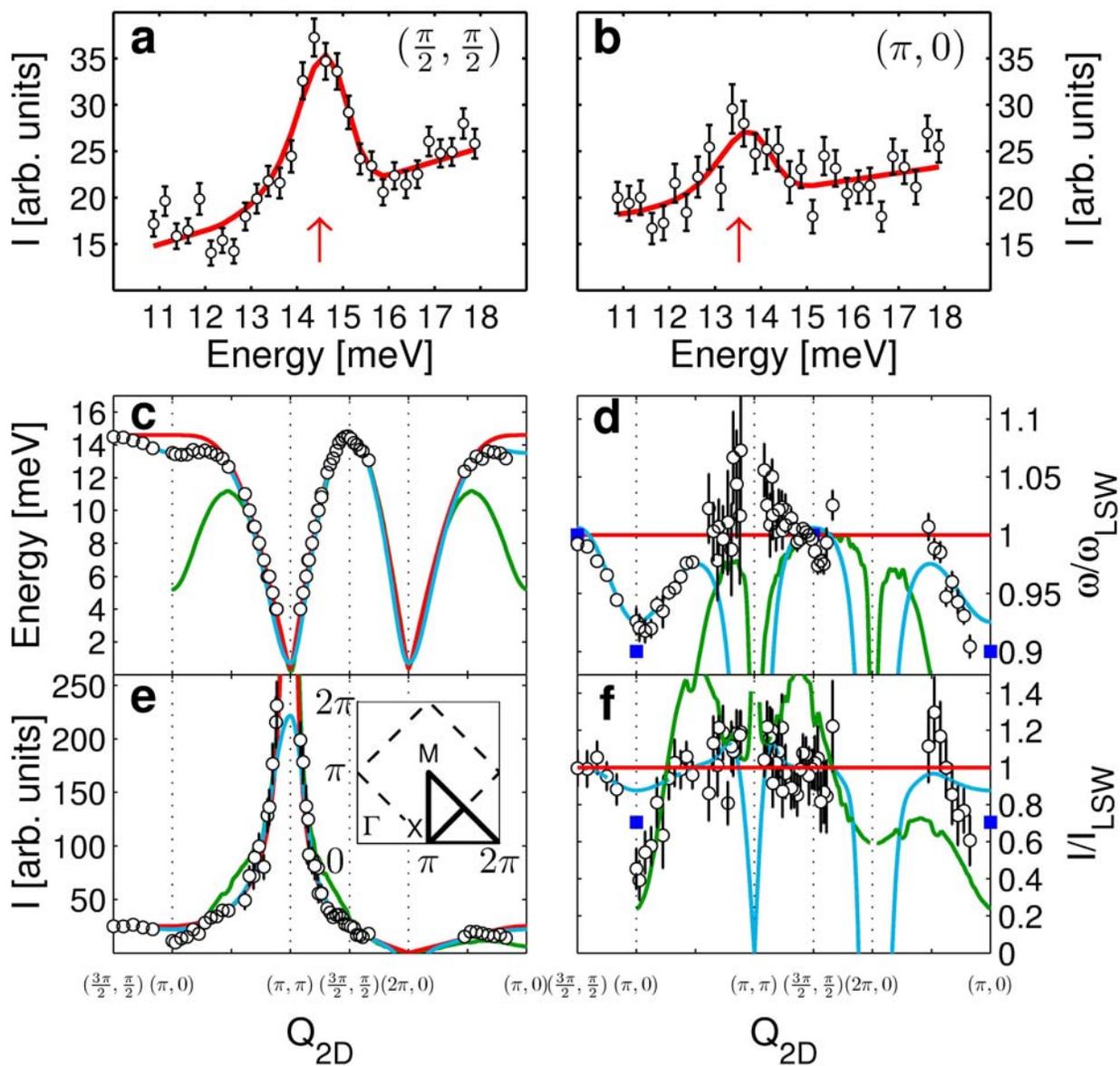

# Figure 4

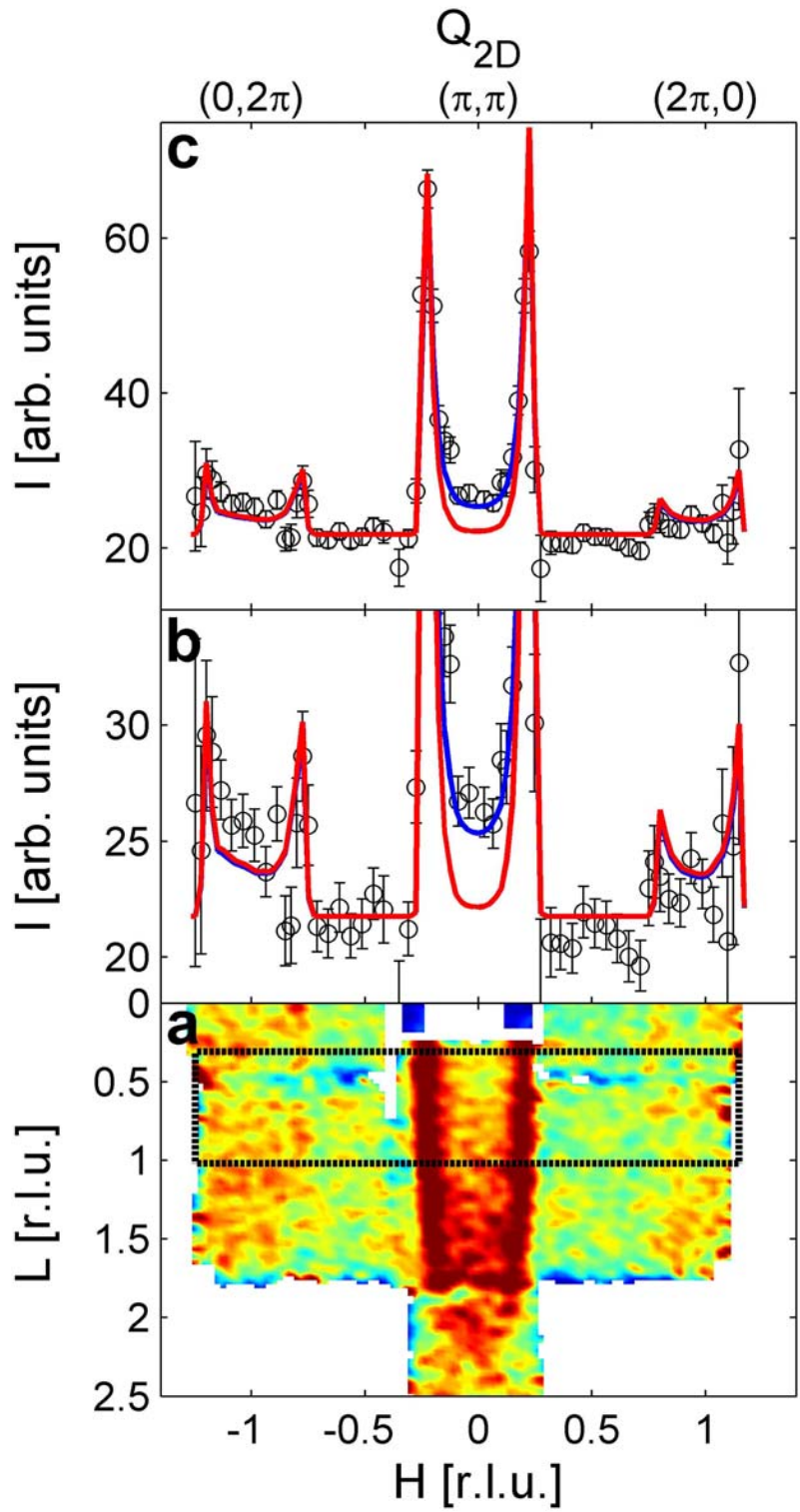

**Figure 5**

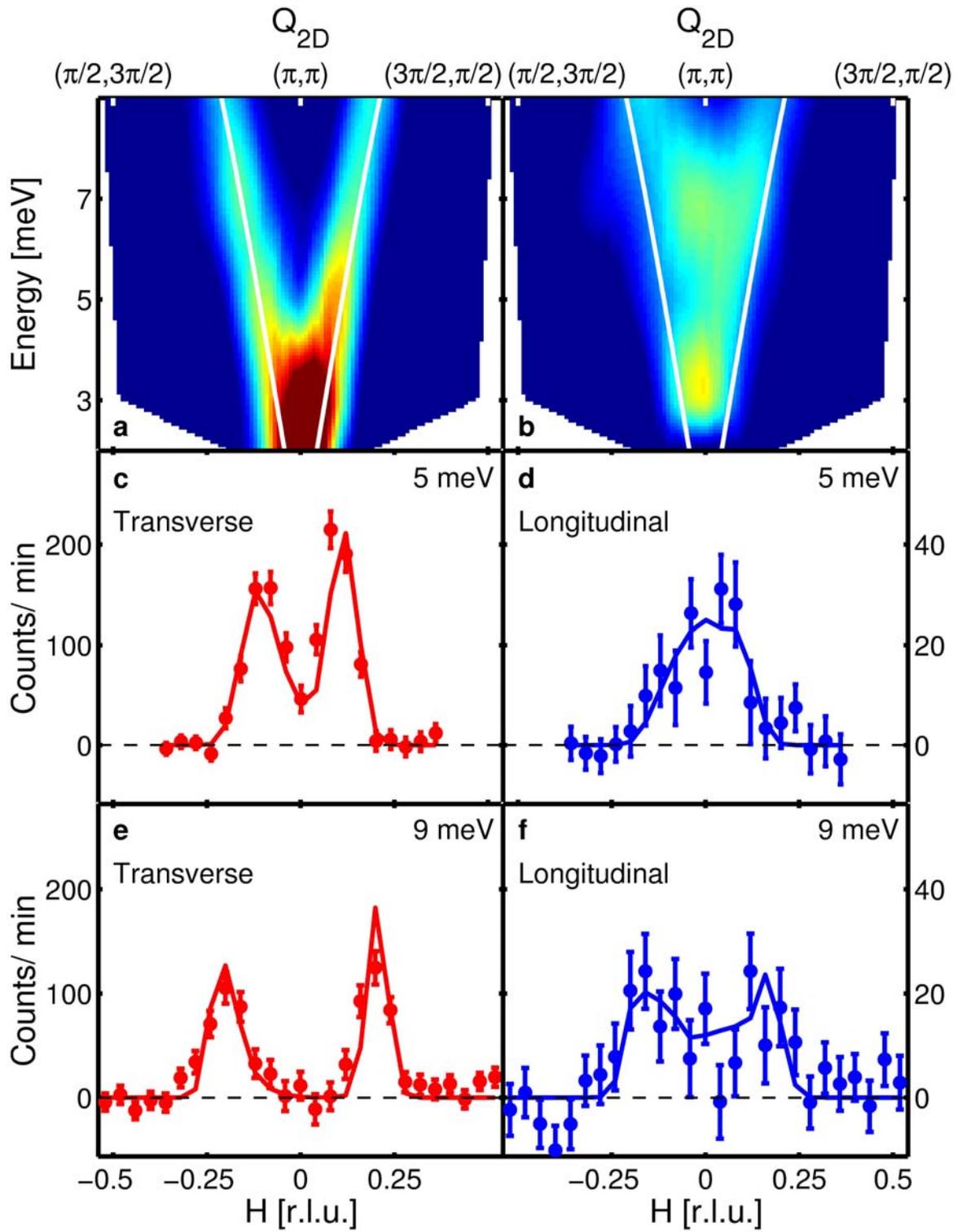

# Figure 6

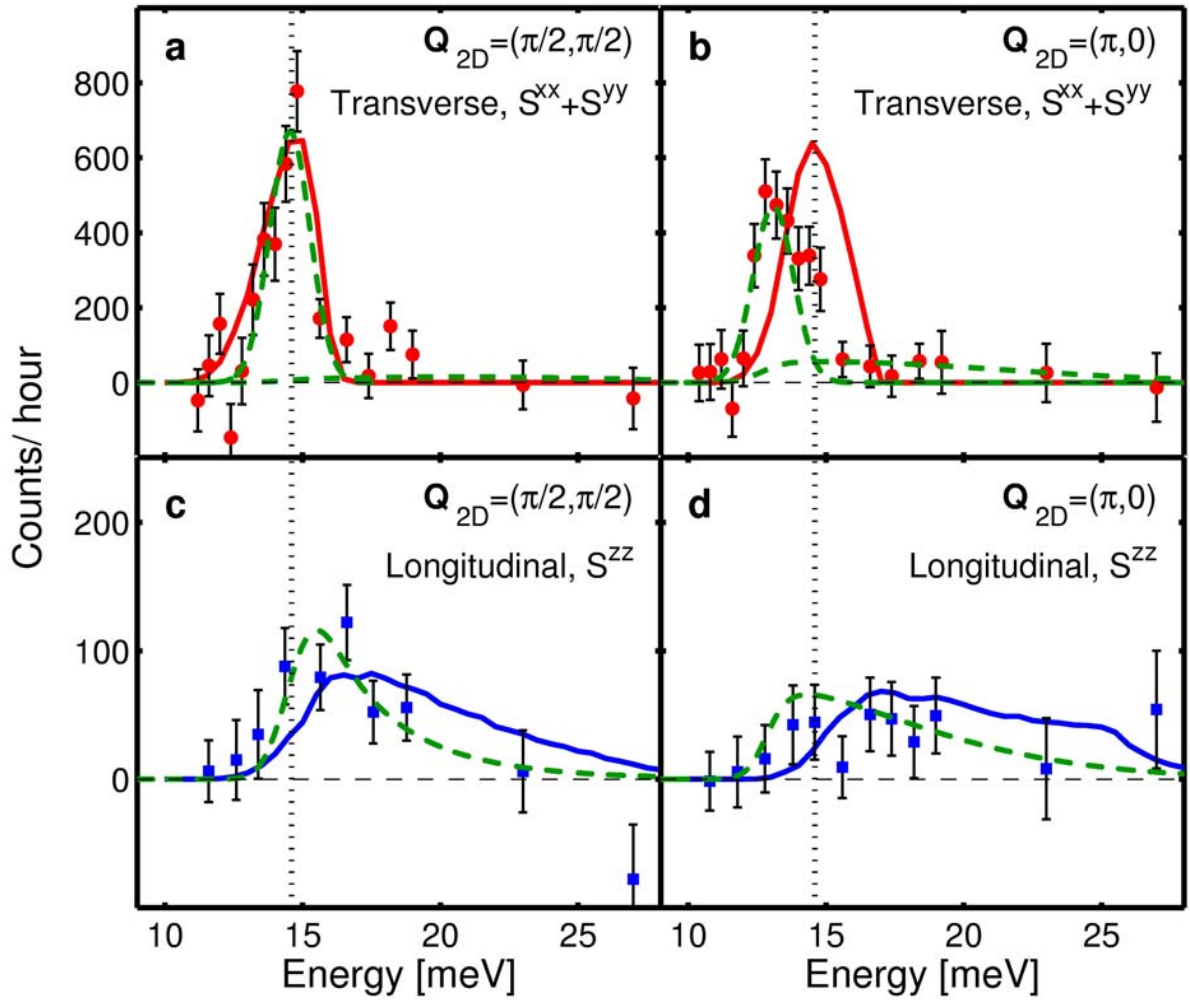